\pdfoutput=1
\documentclass{article}
\RequirePackage[numbers, authoryear]{natbib}
\usepackage{tablefootnote} 
\usepackage{threeparttable}
\usepackage{comment}
\usepackage{draftcopy}
\usepackage{graphicx}%
\usepackage{multirow}%
\usepackage{amsmath,amssymb,amsfonts}%
\usepackage{amsthm}%
\usepackage{mathrsfs}%
\usepackage[title]{appendix}%
\usepackage{xcolor}%
\usepackage{textcomp}%
\usepackage{manyfoot}%
\usepackage{booktabs}%
\usepackage{algorithm}%
\usepackage{algorithmicx}%
\usepackage{algpseudocode}%
\usepackage{listings}%
\usepackage{enumerate}
\usepackage[colorlinks=true, citecolor=black, urlcolor=blue]{hyperref}
\usepackage{authblk}
\usepackage{placeins}
\usepackage{float}
\usepackage{subcaption}

\usepackage{multicol}

\providecommand{\keywords}[1]
{
  \small	
  \textbf{\textit{Keywords:}} #1
}

\begin{document}
\title{Stochastic Claims Reserving Using State Space Modeling}
\author{Rajesh Selukar}
\affil{SAS Institute Inc., Cary NC \\rajesh.selukar@sas.com} 
\date{}
\maketitle
\begin{abstract}
Claims reserving, also known as Incurred But Not Reported (IBNR) claims prediction, is an important issue in general insurance. State space modeling is widely recognized as a statistically robust method for addressing this problem. In state space model-based claims reserving, the Kalman filter and Kalman smoother algorithms are employed for model fitting, diagnostics, and deriving reserve estimates. Additionally, the simulation smoother algorithm is used to obtain the sampling distribution of the derived reserve estimate. The integration of these three algorithms results in an elegant and transparent claims reserving process.

Various state space models (SSMs) have been proposed in the literature for claims reserving. This article outlines a step-by-step process for computing the SSM-based reserve estimate and its associated sampling distribution for any proposed SSM. A brief discussion on model selection is also included. The claims reserving computations are demonstrated using a real-life data set.  The state space modeling computations in the illustrations are performed by using the CSSM procedure in {SAS Viya$^\circledR$}/Econometrics software. The SAS code for reproducing the output in the illustrations is provided in the supplementary material.
\end{abstract}
\keywords{Claims Reserving, IBNR, Simulation Smoother, State Space Model}
\section{Introduction}
\begin{figure}[h]
\caption{Claims Triangle}
\label{tab1}
\begin{center}
\includegraphics[scale=0.60]{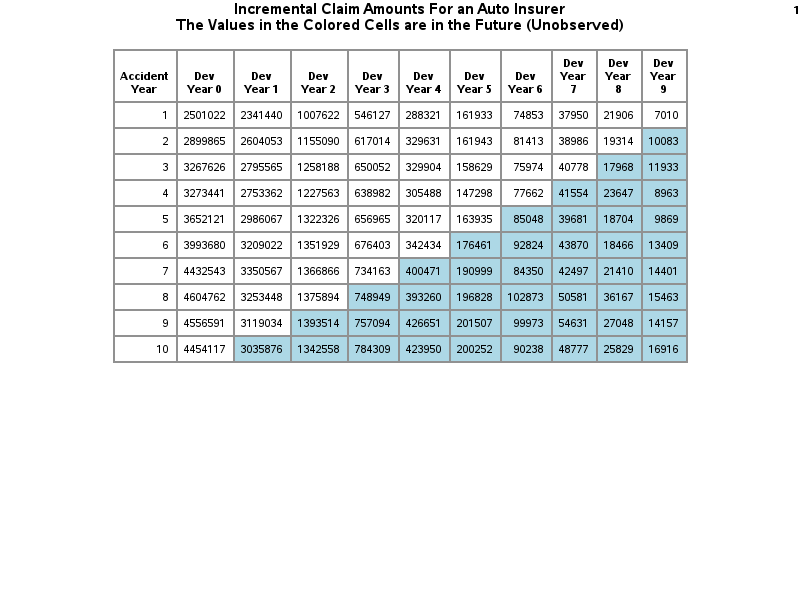}
\end{center}
\end{figure}
The table in Figure~\ref{tab1} shows yearly claims paid by an auto-insurer 
over a stretch of 10 years (to
ensure privacy, this table is created by making minor changes to a real historical claims table).  For accidents
that occur in a given year (accident year), claims are made by the policy holders in that and 
subsequent years, which are called development years for that accident year.  
  The individual cells of the table, $x_{ij}$,
denote the total claims paid by the insurer for an accident year $i$ and development year $j$ 
($ i=1, \cdots, 10$ and $ j=0, \cdots, 9$).
Note that, by the 10th accident year only the upper left triangle of the table is filled with the incurred claims.
The remaining part of the table (the shaded lower right triangle) is unobserved and is filled in 
the subsequent 9 years.
In order to comply with the regulatory guidelines and their own
business needs, the insurers must reserve adequate capital for satisfying the total future 
claims (which is
the sum of $x_{ij}$ in the shaded triangle).  This is the well-known claims-reserving problem.
There are many variants of this problem, for example, the historical information may be more or less,
the forecast horizon may be longer or shorter, and the time intervals could be different.
Nevertheless, the following aspects of this problem remain the same:
\begin{itemize}
\item the reserve amount must be predicted based on a relatively small amount of data (e.g.,
only the data from the 55 cells in the upper left triangle (the unshaded part) of Figure~\ref{tab1} are available
to predict the sum of future claims in the  45 cells in the lower right triangle (the shaded part)).
\item since the future is uncertain and the cost of under/over-reservation can be high, a point estimate of
the reserve is not enough.  The sampling distribution of the reserve estimate is needed.
\item since the insurance companies work in regulated environments, the reserving process
must be interpretable, transparent, and statistically sound.
\end{itemize}
These aspects make the claims reserving problem a challenging one.  
Many methods, some that are based on statistical modeling and others that are more
heuristic, are in use.   The methods that are based on statistical modeling are called
stochastic claims reserving methods.  There are many stochastic claims reserving methods
based on a variety of statistical models such as GLMs (generalized linear models),
SSMs, and based on Bayesian and frequentist approaches.  The review of all these methods
is beyond the scope of this article.  You can get a flavor of GLM-based methods from 
\cite{Taylor16}, see \cite{CL} for some well-known heuristic and stochastic methods, 
 and for a review of SSM-based claims reserving methods, see \cite{Chukhrova}.

This article deals with SSM-based claims reserving, which has a long history;
for example, see \cite{deJong83, verr94, deJong04, ather, Hendrych, Chukhrova}.
The SSMs are well suited for the claims reserving problem because
\begin{itemize}
\item it is a large and flexible class of models that naturally
incorporates the longitudinal nature of the claims process.
\item the models can be customized to take into account a variety of claims patterns, 
special business logic, and outliers in the data.
\item it is a mature model-class with well-understood process for model fitting, model diagnostics and
comparison, forecasting and interpolation of the claims process, and the generation of 
the sampling distribution of the claims-reserve estimate.
\end{itemize}
In SSM-based modeling, the model fitting (parameter estimation), model diagnostics, and the point estimation 
of claims-reserve is carried out by using the well-known Kalman filter 
and Kalman smoother algorithms.  These two algorithms are widely known,
 however, for the important problem of obtaining the sampling distribution of the reserve 
estimate, a third algorithm, the simulation smoother, is needed.
While well-known in other fields, the use of simulation smoother for obtaining the 
sampling distribution of claims-reserve is relatively new, see \cite{Hendrych}.
Much earlier, without mentioning the simulation smoother explicitly, 
simulation smoothing-based sampling distribution of the reserve estimate
is advocated in \cite{deJong04}.

The aim of this article is to provide a step-by-step recipe for SSM-based claims-reserving
process.  The article is organized as follows:
\begin{itemize}
\item Subsections of Section~\ref{sec1} provide a step-by-step process for computing
the SSM-based reserve estimate and its associated sampling distribution for any
proposed SSM.
\item Section~\ref{Summary} summarizes the article and outlines the plans for future work.
\item Appendix~\ref{SSMFramework} provides the background
and references for the state space modeling framework used in this article.
\end{itemize}

The state space modeling computations in the illustrations in this article are
done by using the CSSM procedure in {SAS Viya$^\circledR$}/Econometrics software
(for more information, see \cite{cssm1}).  Even though it is generally agreed that stochastic 
reserving methods are
to be preferred because of their statistical transparency, Chain-Ladder (CL), a heuristic method, 
 is by far the most widely used reserving method in practice.
Therefore, we will use the CL method as the benchmark method in these 
illustrations.   The R package \emph{ChainLadder} (see \cite{CL}) is 
used to calculate the CL-based reserve estimate and it's standard error (which 
is calculated by a method described in \cite{mack93}).

\section{SSM-Based Claims Reserving}
\label{sec1}
The entries, $x_{ij}$, in the table in Figure~\ref{tab1} 
are called incremental claims.  For modeling purposes, it is useful to consider the same 
information in a few alternate forms:
\begin{itemize}
\item Logarithm of incremental claims: $\log(x_{ij})$
\item Cumulative claims within accident years (cumulative row sums): $C_{ij} = \sum_{k=0}^{j} x_{ik}$
\item Logarithm of cumulative claims within accident years: $\log(C_{ij})$
\item Development ratios across accident years: $D_{i0}=C_{i0}$ and for $j = 1, 2, \cdots,$ $D_{ij} = C_{ij}/C_{i(j-1)}$
\item Logarithm of development ratios across accident years: $\log(D_{ij})$
\end{itemize}
Information content in these alternate forms is the same, i.e., the numbers in one form
can always be converted to any other form.  
When there are no data irregularities, the
incremental claims, $x_{ij}$, are nonnegative (and so are $C_{ij}$ and $D_{ij}$).  
Therefore, SSM-based reserving algorithms often work with
their log-transformed versions ($\log(x_{ij})$, $\log(C_{ij})$, and $\log(D_{ij})$),
which helps with the assumption of Guassianity of the response variable 
that underpins these methods. 
Apart from using different numeric forms, different SSM-based reserving algorithms 
process the numbers in the claims table in different sequence, e.g.,
some algorithms process the numbers by the development years (row-wise) and some
others process them by calendar years.  The calendar year processing corresponds to 
the way the claims are naturally reported over the years and correspond to  
lower-left to upper-right diagonals of the claims table (i.e., for a calendar year $t$, the
row and column indices of the entries in the diagonal satisfy the relation 
$t = (i+j), i=1, 2, \cdots, j=0, 1, \cdots$).  Table~\ref{Models1} summarizes this information for the
SSMs that are used in the illustrations in this article. 
\begin{table*}[h]
\begin{center}
\caption{Response Variable and Sequence Type for Some SSMs}
\label{Models1}
\begin{tabular}{@{}c c c c @{}}
\hline
\multicolumn{1}{c}{Model Name} &   \multicolumn{1}{c}{Reference}  & 
\multicolumn{1}{c}{Response Variable} &
\multicolumn{1}{c}{Sequence}  \\
\hline
BSM    &   \cite{ ather}  &  Log(Incremental claims) &  Row-Wise \\  
CC    &   \cite{ deJong04}  &  Log(Development ratios) &  Calendar year \\ 
Hertig    &   \cite{ deJong04}  &  Log(Development ratios) &  Calendar year \\
Verral    &   \cite{ verr94}  &  Log(Incremental claims) &  Calendar year \\
\hline
\end{tabular}
\end{center}
\end{table*}

Of course, no matter what response variable is used by the reserving algorithm or 
in which sequence it processes the claims information, the ultimate goal is to
obtain a point estimate, $\hat{R}$, of the claims reserve
$R = \sum  x_{ij}$ (the sum is over $x_{ij}$ in the unobserved lower triangle),
and the sampling distribution of $\hat{R}$.  
In the remainder of this section we describe the main steps an SSM-based reserving  algorithm
follows for achieving this goal.  We will assume that the initial input for all algorithms
is a table of incremental claims, $x_{ij}$, as in Figure~\ref{tab1},
with missing values (NaN) in the shaded region.  In order to simplify
the description, we will illustrate all the steps with the table 
in Figure~\ref{tab1}, which shows yearly claims for a 10 year stretch.  
Essentially the same pattern carries over for tables of other
dimensions and other timing intervals.  The SAS code for reproducing the output in the illustrations
is provided in the supplementary material.

\subsection{Data Transformation and Organization}
\label{subsec2}
\begin{figure}[h]
\caption{Transformed Claims}
\begin{center}
\includegraphics[scale=0.28]{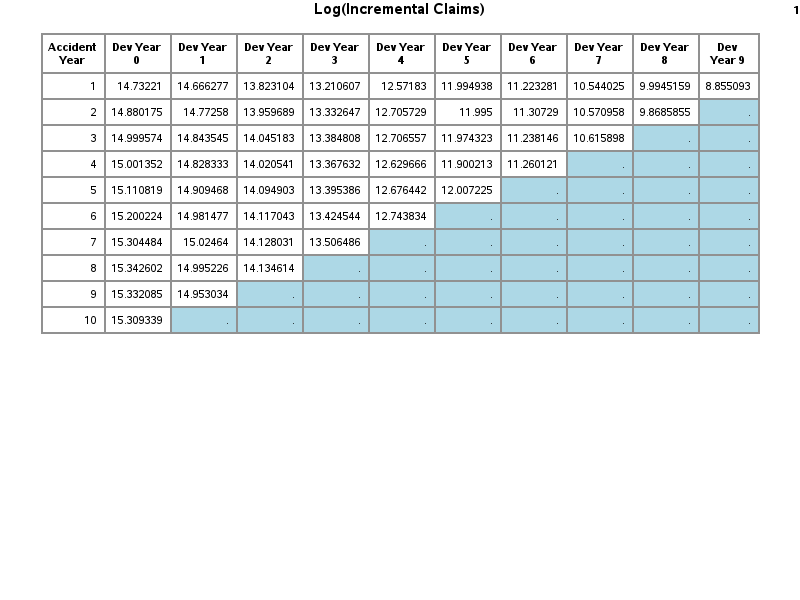}
\includegraphics[scale=0.28]{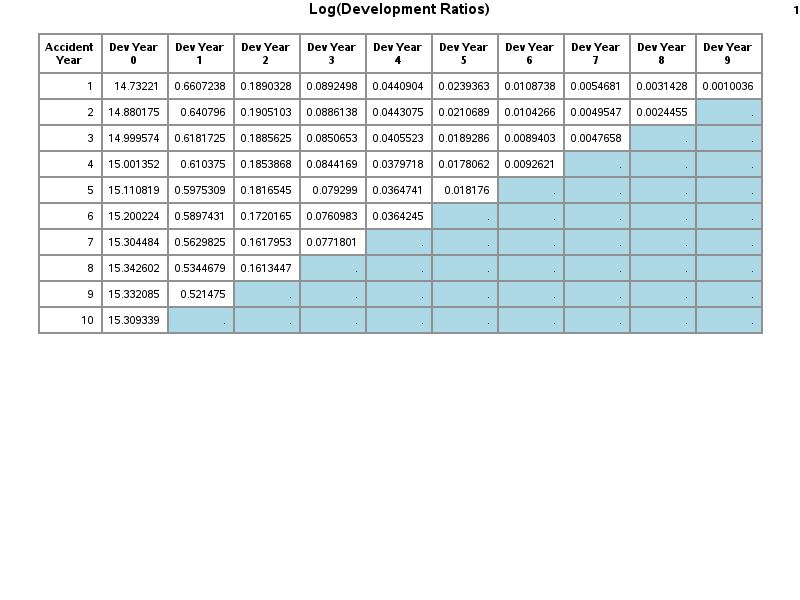}
\end{center}
\label{tab2}
\end{figure}

The first step is to transform the input data appropriately and to
assign a time index to the observations that aligns with the sequential access
needed for the SSM that is used by the reserving  algorithm.
For example, Figure~\ref{tab2} shows two such transformations of the claims values,  
$\log(x_{ij})$ and $\log(D_{ij})$,  
for the table in Figure~\ref{tab1}.
We will denote the transformed values, the 
response variable for the SSM that is used by the reserving  algorithm,
by  $y$.  That is, $y$ will often be either $\log(x_{ij})$, $\log(C_{ij})$, or $\log(D_{ij})$.
If the SSM's sequential access is row-wise, the $y$ values are indexed as
$(y_{t}, t=1, 2, \cdots, 100)$ where the index $t$ starts in the first cell and increases
row-wise, finally reaching up to 100.  This access pattern creates a time series
$y_t$ that has embedded missing values, for example, 
the last value in the second row ($y_{19}$), 
the last two values in the third row ($y_{29}, y_{30}$), 
and the last 9 values in the 10th row ($y_{92},  \cdots, y_{100}$),  
are
missing.   If the SSM's sequential access is calendar-year-wise,
the $y$ values are indexed as $({\bf{y}}_{t}, t=1, 2, \cdots, 19)$ where
the vectors ${\bf{y}}_{t}$ contain the claims from the calendar year
$t$, which correspond to the lower left to upper right diagonals.
Thus, ${\bf{y}}_{1}$ has one value ($y_{1,0}$),  ${\bf{y}}_{2} $ has two
values ($y_{2,0}, y_{1,1}$), and so on.  For calendar years 11 through 19,
 ${\bf{y}}_{t}$ has missing values.

\subsection{Model Fitting, Diagnostics, Forecasting, and Interpolation}
\label{subsec3}
After the data are transformed and indexed, they form a time series
(like $(y_{t}, t=1, 2, \cdots, 100)$ or $({\bf{y}}_{t}, t=1, 2, \cdots, 19)$) that
is ready for modeling by the chosen SSM.   In Appendix~\ref{SSMFramework},
we describe an SSM framework that is general enough to handle
the different types of time series and SSMs that arise in
different claims reserving algorithms.  In particular, this framework permits
\begin{itemize}
\item Different number of observations at different time points.
\item Time-varying system matrices.
\item Partially diffuse initial condition.
\end{itemize}
Once a problem is formulated as an SSM, 
 model fitting, diagnostics,
and forecasting (and interpolation) of the response values is done in a standard way
by using the (diffuse) Kalman filter and Kalman smoother algorithms
(for more details, see Appendix~\ref{SSMFramework}).  For illustration purposes,
we will fit the four models in Table~\ref{Models1} to the appropriately transformed 
(and indexed) data in Figure~\ref{tab2}.  These models are just a small sample from
a large variety of SSMs that are available for the modeling of claims reserves.
Nevertheless, this illustration will highlight several important issues
a modeler must consider.

The CSSM procedure in {SAS Viya$^\circledR$}/Econometrics software that is used here
for state space modeling provides a large variety of output that includes
\begin{itemize}
\item (marginal) maximum likelihood  estimates of model parameters.
\item marginal likelihood-based information criteria such as AIC and BIC.
\item model diagnostics based on one-step-ahead residuals as well as delete-one
cross validation. 
\item detection of additive outliers and structural breaks.
\item forecasts and interpolations of response values and the latent components
in the model.  
\end{itemize}
This output is based on Kalman filter and Kalman smoother.  
After model fitting and forecasting, you can 
obtain a point estimate of $R$, which we will denote by $\hat{R}$, after appropriate inverse transformation 
and aggregation of the forecasted
(or interpolated) response values.  However, in order to obtain the sampling distribution of
$\hat{R}$, simulation smoother must be used. How to  
obtain the simulation smoother-based sampling distribution of $\hat{R}$ is described in 
Subsection~\ref{subsec4}.

\begin{figure}[h]
\caption{Model-Based Predictions}
\begin{center}
y = Log(Development Ratios) \\
\includegraphics[scale=0.28]{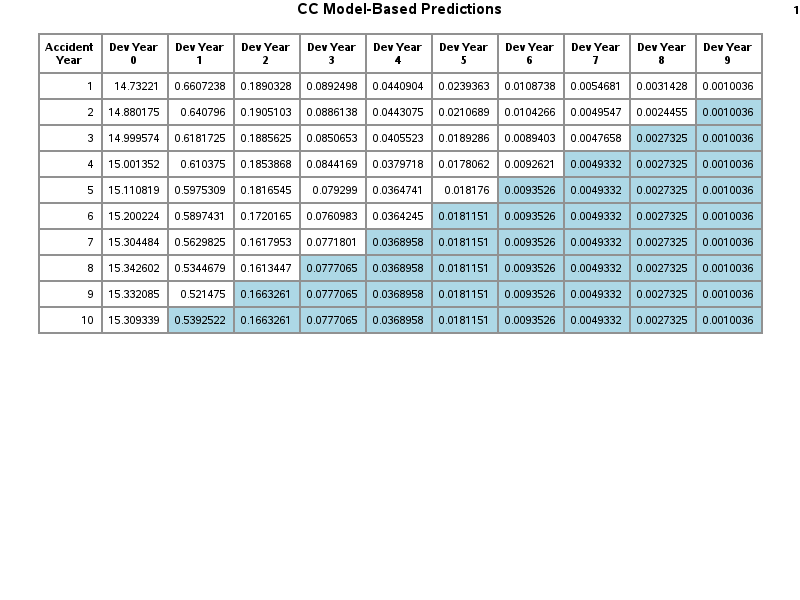}
\includegraphics[scale=0.28]{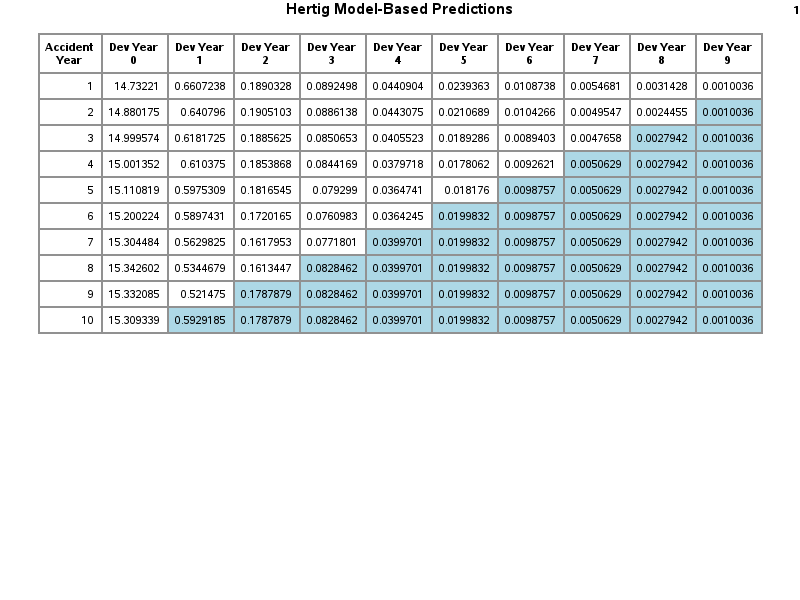}
\\
y = Log(Incremental Claims)\\
\includegraphics[scale=0.28]{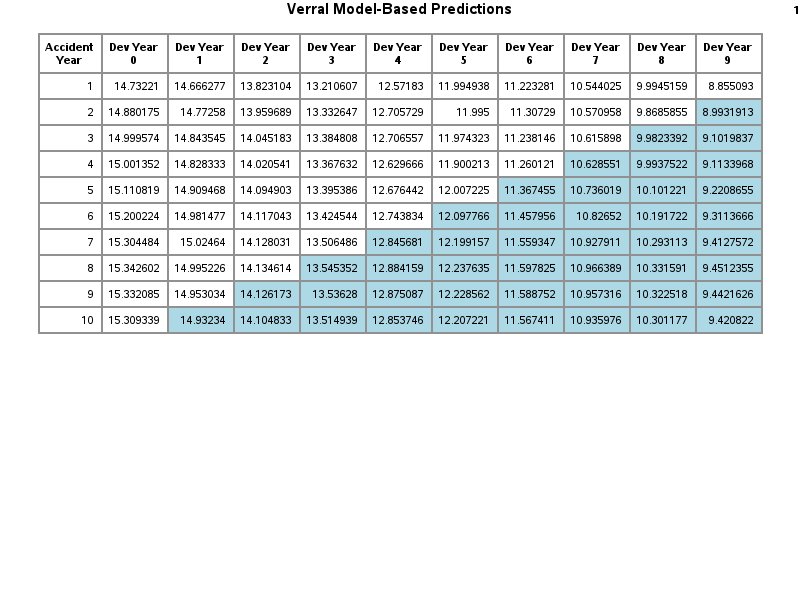}
\includegraphics[scale=0.28]{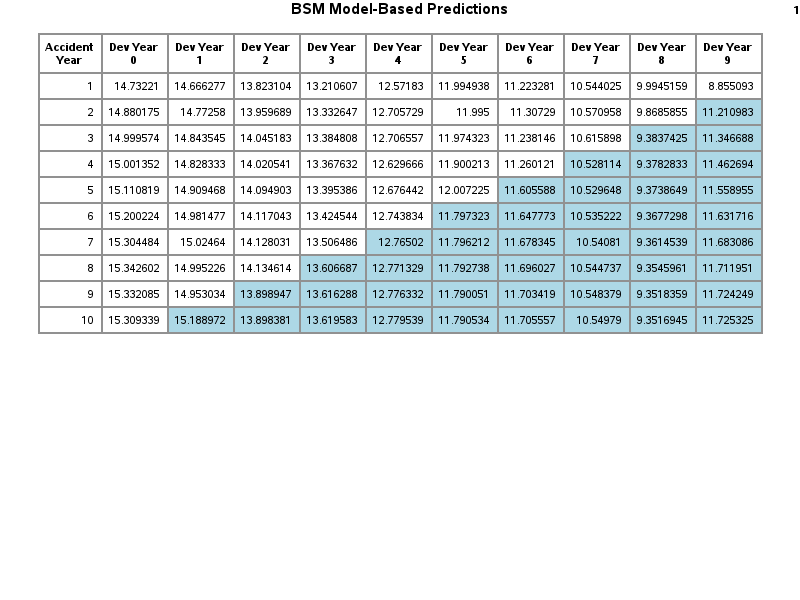}
\end{center}
\label{tab3}
\end{figure}

In the remainder of this subsection we briefly review the output of this
fitting-diagnostics-forecasting phase.  For brevity sake, we  
consider only two parts of this output: the model comparison on the basis of BIC 
(a popular likelihood-based information criterion), and the prediction/interpolation of the
response values.
Starting with the model comparison, Table~\ref{InfoLogDR} shows the BIC values
(in smaller-is-better form) for the four models.  It
divides the models according the response variable because the BIC values are 
comparable only when the models have the same response variable.
Based on this table, when the response variable is $Log(Development Ratios)$,
the CC model is  preferred over the
Hertig model and when the response variable is $Log(Incremental Claims)$,
the Verral model is preferred over the BSM model.  Finally, the tables in 
 Figure~\ref{tab3} show the predictions
of the future  response values,  $\hat{y}$, according to these models (predicted
values are in the shaded region).  These predicted values will have to be inverse
transformed to obtain the predicted future incremental claims ($\hat{x}_{ij}$), which are
then aggregated to obtain a point estimate of the claims-reserve, $\hat{R}$. 
\begin{table*}[h]
\begin{center}
\caption{BIC Information Criterion for Different Models}
\label{InfoLogDR}
\begin{tabular}{@{}c c c c @{}}
\hline
\multicolumn{2}{c}{y = Log(Development Ratios) }& 
\multicolumn{2}{c}{y = Log(Incremental Claims) } \\
\hline
\multicolumn{1}{c}{Model} &   \multicolumn{1}{c}{BIC}  & 
\multicolumn{1}{c}{Model} &
\multicolumn{1}{c}{BIC}  \\
\hline
CC         &  -248.9  &  Verral  &  -100.1  \\  
Hertig   &  -232.4  &  BSM &   72.0 \\ 
\hline
\end{tabular}
\end{center}
\end{table*}

\subsection{Point Estimate of $R$ and its Sampling Distribution}
\label{subsec4}
\begin{figure}[h]
\caption{Inverse Transformed Predictions}
\begin{center}
\includegraphics[scale=0.28]{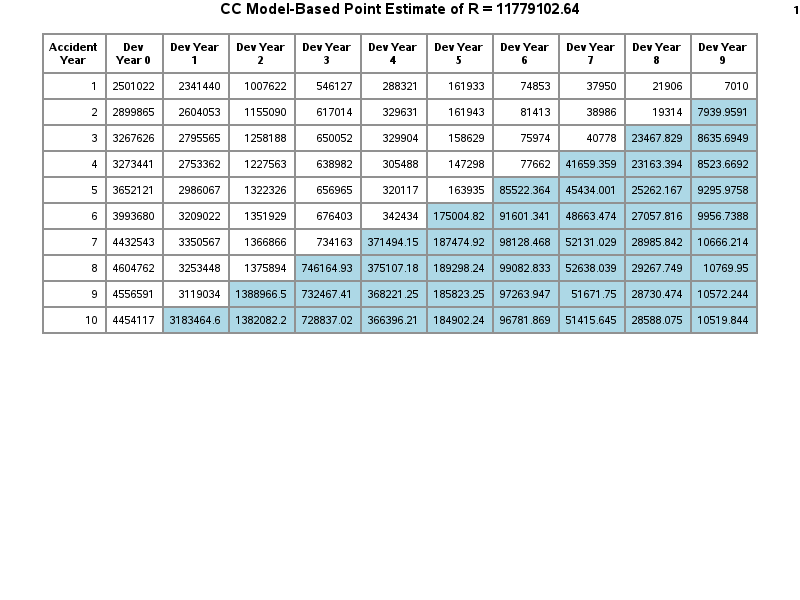}
\includegraphics[scale=0.28]{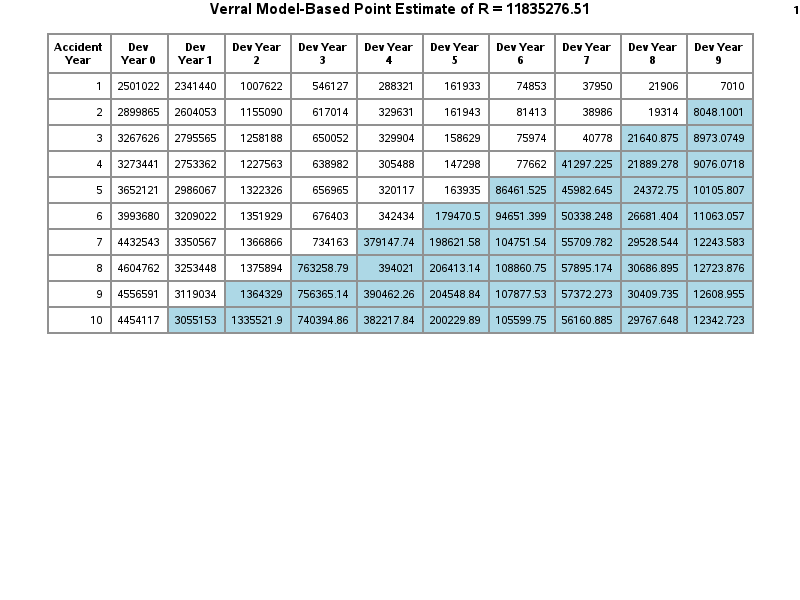}
\includegraphics[scale=0.28]{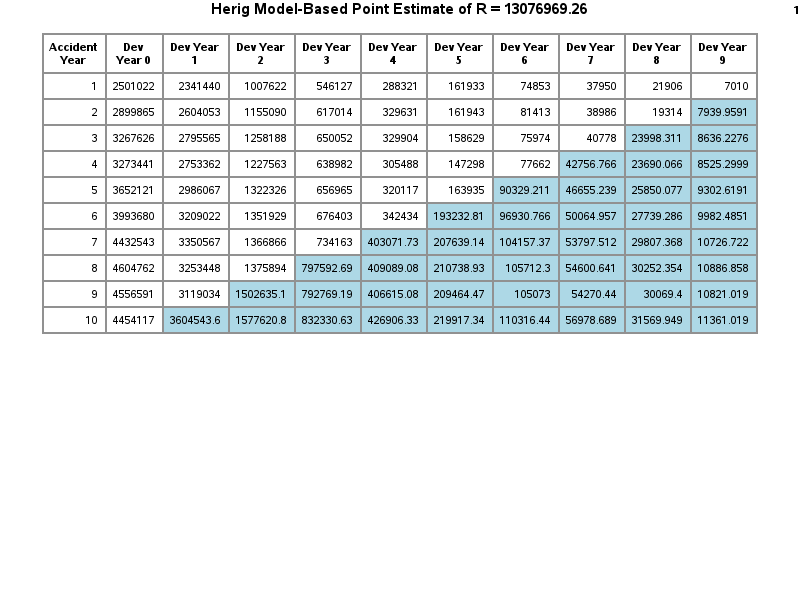}
\includegraphics[scale=0.28]{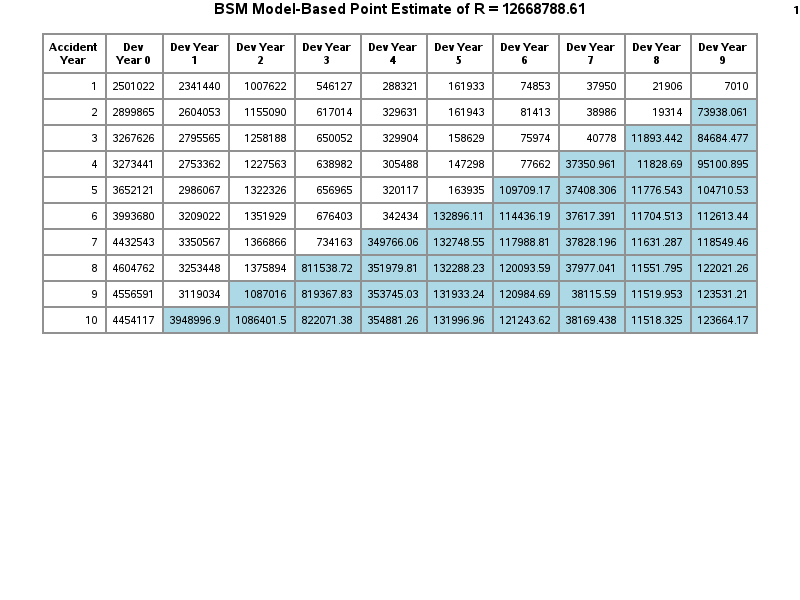}
\end{center}
\label{tab4}
\end{figure}

In this subsection we will show how to obtain the sampling distribution
of the reserve estimate, $\hat{R}$, which completes the solution of the claims reserving
problem.  As a first step, we see how to obtain $\hat{R}$ by 
inverse transforming the model predictions, $\hat{y}$, that we saw in Figure~\ref{tab3}.
These inverse transformed predictions, $\hat{x}_{ij}$,  are shown in Figure~\ref{tab4}.
For each table in Figure~\ref{tab4}, $\hat{R}$ is obtained by summing $\hat{x}_{ij}$,
the values in the shaded region.
Note that for the claims table used
in our illustration we know both the historical and future incremental claims
(the shaded area in the table in Figure~\ref{tab1}), and therefore
the exact total of the incurred claims in the future--the ground truth--is 
also known, which is 11,854,009 (about 11.85 million).  
Additionally,
the point estimate provided by the ChainLadder method for this table turns out to be 
12905462.98 (about 12.90 million).    All this information is summarized in Table~\ref{pointR}.  
\begin{table}[h]
\caption{Point Estimates of R by Different Methods (in Millions)}
\label{pointR}
\begin{center}
\begin{tabular}{@{}c c c c c @{}}
\multicolumn{5}{c}{Known Value of R = 11.85} \\
\hline
\multicolumn{1}{c}{CC} &   \multicolumn{1}{c}{Verral}  & 
\multicolumn{1}{c}{Hertig} &
\multicolumn{1}{c}{BSM}  &
\multicolumn{1}{c}{ChainLadder} \\
\hline
11.78  & 11.83 & 13.08 & 12.67 & 12.90 \\
\hline
\end{tabular}
\end{center}
\end{table}
From this summary it appears that the historical claims patterns
in the first 10 years continued in the subsequent 9 years and
the CC and Verral models that fit the historical data well, 
at least according to the BIC criterion,
predicted $R$ better than the other models.  Among the four models we have considered
in this illustration, the Hertig model is closest in spirit to the
assumptions that underlie the ChainLadder method and it is not
surprising that the point estimates based on the Hertig model
and the ChainLadder method are somewhat close.
The Hertig model is a pure regression model with time-invariant regression
effects, while the other three models (BSM, CC, and Verral)
can be considered as regression models with time-varying
regression effects.  Having obtained $\hat{R}$, 
we now proceed to the process of obtaining the sampling distribution of $\hat{R}$.

Like the Kalman filter and Kalman smoother algorithms,
the simulation smoother is an important algorithm in SSM-based data analysis.
The KF and KS algorithms provide the conditional distribution of latent states $\boldsymbol{\alpha}_{t}$
at individual time points $t$, conditional on the observed data 
(based on the partial sample $\mathbf{Y}_{t} = (\mathbf{y}_{s}, s=1, 2, \cdots, t)$ 
in the case of KF, and the full sample
$\mathbf{Y} = (\mathbf{y}_{t}, t=1, 2, \cdots, n) $ in the case of KS).
Such conditional distributions are sufficient for many commonly needed tasks such as 
likelihood computation and prediction/interpolation of response values and latent states at
individual time points.  However, because KF and KS don't provide
the  joint conditional distribution of 
$(\boldsymbol{\alpha}_{1}, \boldsymbol{\alpha}_{2}, \cdots, \boldsymbol{\alpha}_{n})$
 given $\mathbf{Y}_{t}$ or $\mathbf{Y}$, they cannot be used to make statements
about the functions of the entire set of latent vectors
$(\boldsymbol{\alpha}_{1}, \boldsymbol{\alpha}_{2}, \cdots, \boldsymbol{\alpha}_{n})$,
such as the sum of response-variable predictions  or the sum of  
inverse transformed predictions.  The simulation smoother is useful
in precisely these situations because it enables random drawings from the
joint conditional distribution of
 $(\boldsymbol{\alpha}_{1}, \boldsymbol{\alpha}_{2}, \cdots, \boldsymbol{\alpha}_{n})$,
given the observed sample $\mathbf{Y}$.  This enables the computation of the sampling
distribution of any arbitrarily complex function of the conditional estimate of
$(\boldsymbol{\alpha}_{1}, \boldsymbol{\alpha}_{2}, \cdots, \boldsymbol{\alpha}_{n})$.
Therefore, since  $\hat{R}$ is a function of the conditional
estimates of the latent states 
$(\boldsymbol{\alpha}_{1}, \boldsymbol{\alpha}_{2}, \cdots, \boldsymbol{\alpha}_{n})$, 
the simulation smoother enables
you to get random draws of $\hat{R}$  from the
joint conditional distribution of
 $(\boldsymbol{\alpha}_{1}, \boldsymbol{\alpha}_{2}, \cdots, \boldsymbol{\alpha}_{n})$,
given the observed sample $\mathbf{Y}$.
In addition to the model fitting, diagnostics, and forecasting for SSMs,
the CSSM procedure enables you to obtain random draws of
latent states as well as predictions of the response variable from
the conditional distribution of 
$(\boldsymbol{\alpha}_{1}, \boldsymbol{\alpha}_{2}, \cdots, \boldsymbol{\alpha}_{n})$,
given the observed sample $\mathbf{Y}$.   After inverse transformation and aggregation
of the random draws of response variable predictions, you obtain the sampling
distribution of $\hat{R}$.  For a fully self-contained and short example of
the use of simulation smoother for deriving the sampling distribution of
a function of SSM-based predictions, see \url{https://go.documentation.sas.com/doc/en/pgmsascdc/v_061/casecon/casecon_cssm_examples21.htm}.

\begin{figure}[!h]
\caption{Summary Measures for the Sampling Distribution of $\hat{R}$}
\label{tab6}
\centering
\begin{subfigure}{\textwidth}
\includegraphics[width=\textwidth]{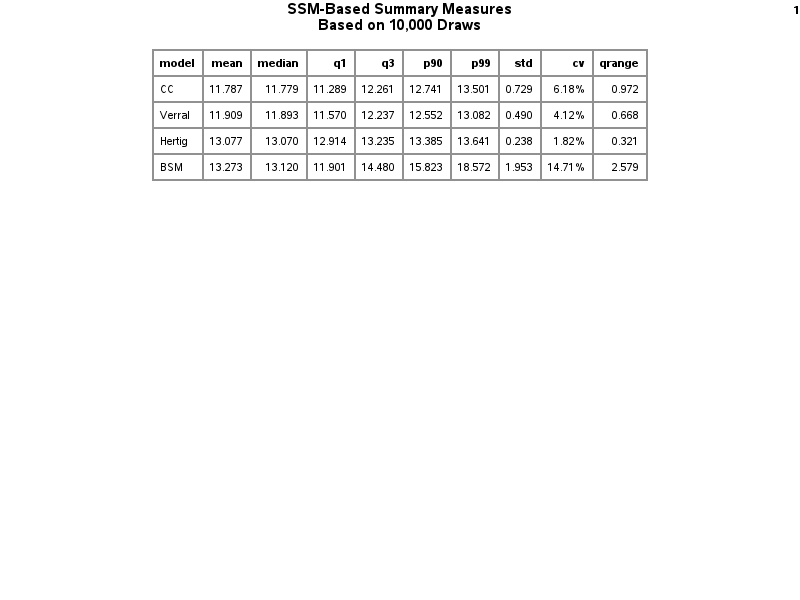}
\end{subfigure}
\end{figure}

In our illustration, we have computed the sampling distributions of $\hat{R}$ based on 10,000 random draws
of $\hat{R}$ from the conditional distribution.  For additional accuracy, you can
increase the number of draws (of course, with increased computational burden).  We
start the exploration of these sampling distributions by calculating some basic
summary measures, which are shown in Figure~\ref{tab6}.  For comparison, Table~\ref{chainSum}
shows the summary measures based on the ChainLadder method.
For improved readability, all summary measures, except the coefficient of variation (CV), 
are in the units of millions.  
\begin{table}[h]
\caption{Summary Measures for the Chain Ladder Method}
\label{chainSum}
\begin{center}
\begin{tabular}{@{}c c c c c@{}}
\hline
 \multicolumn{1}{c}{CL Reserve}  & 
\multicolumn{1}{c}{CL StdError} &
\multicolumn{1}{c}{Reserve+SE}  &
\multicolumn{1}{c}{Reserve+2SE}  &
\multicolumn{1}{c}{CV} \\
\hline
12.905 & 0.564& 13.469& 14.033 & 4.37\% \\
\hline
\end{tabular}
\end{center}
\end{table}

\begin{figure}[!h]
\caption{ $\hat{R}$ Sampling Distributions (with Differently Scaled X-Axis)}
\begin{center}
\includegraphics[scale=0.28]{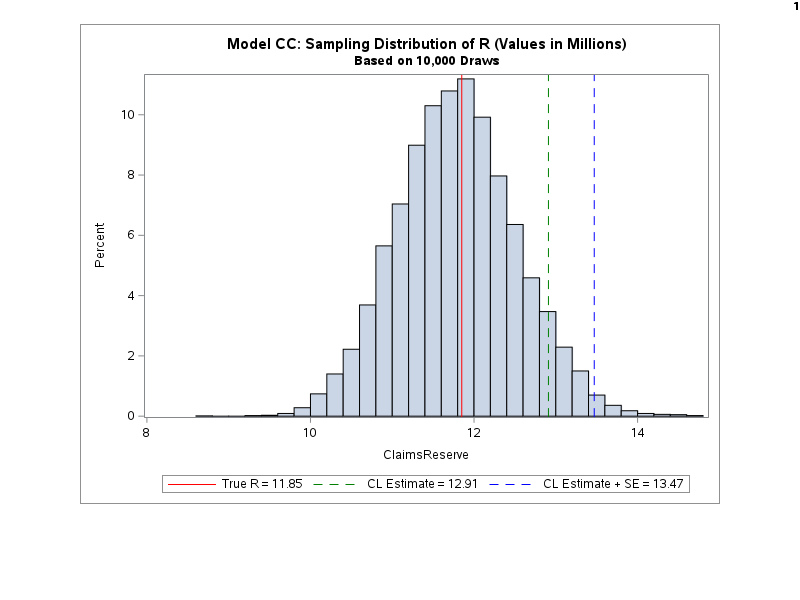}
\includegraphics[scale=0.28]{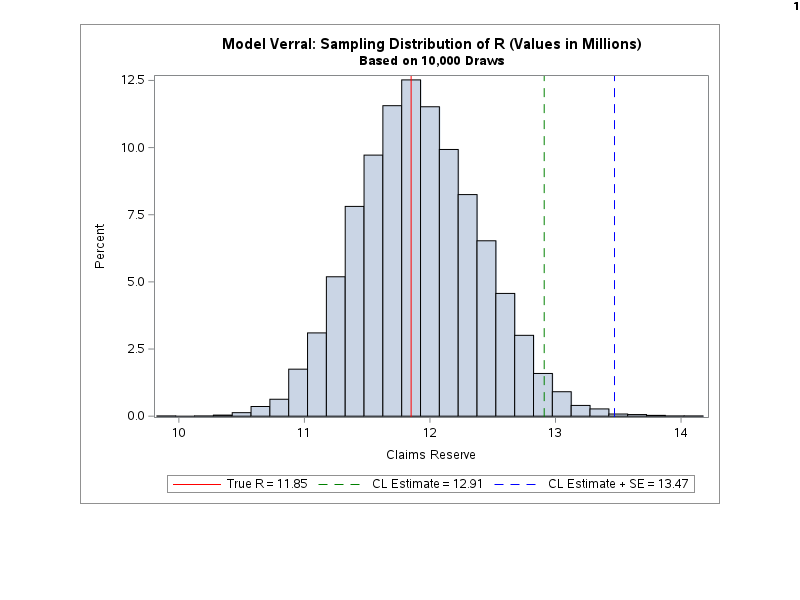}
\includegraphics[scale=0.28]{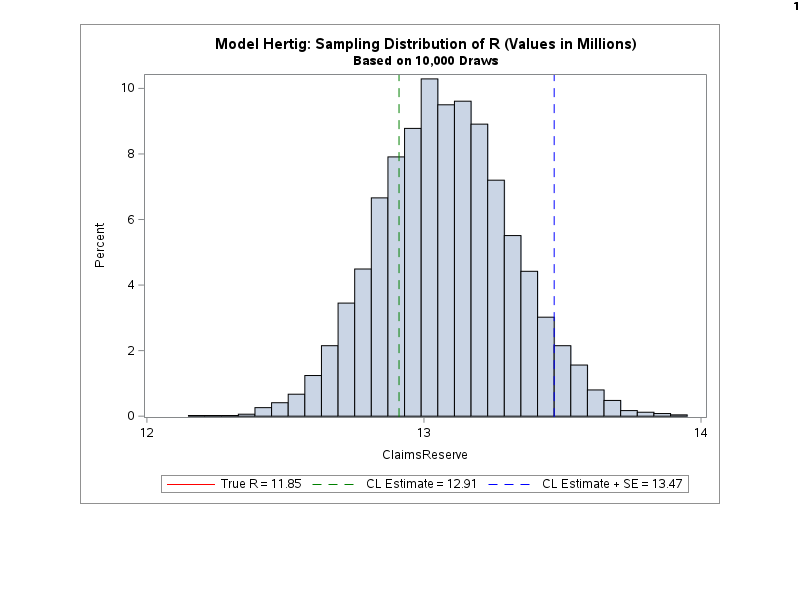}
\includegraphics[scale=0.28]{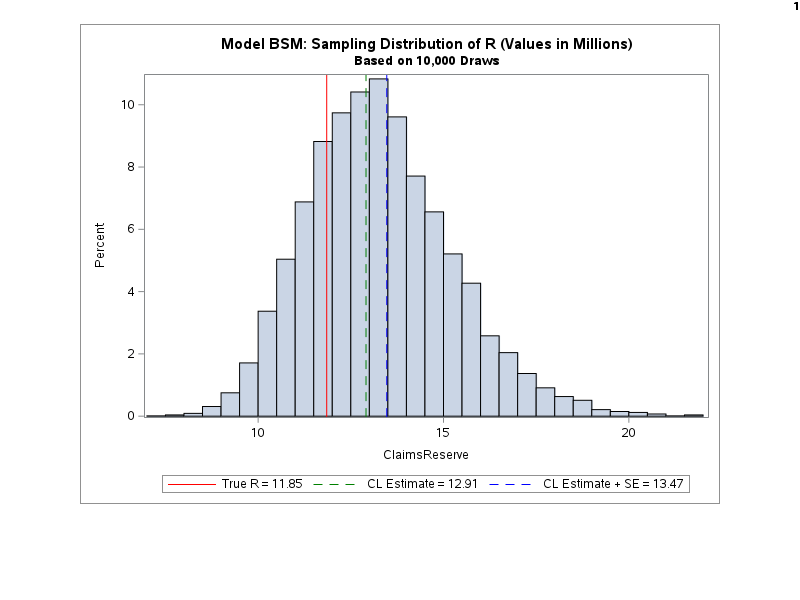}
\end{center}
\label{tab7}
\end{figure}

\begin{figure}[!h]
\caption{ $\hat{R}$ Sampling Distributions (with the Same Scaled X-Axis)}
\begin{center}
\includegraphics[scale=0.65]{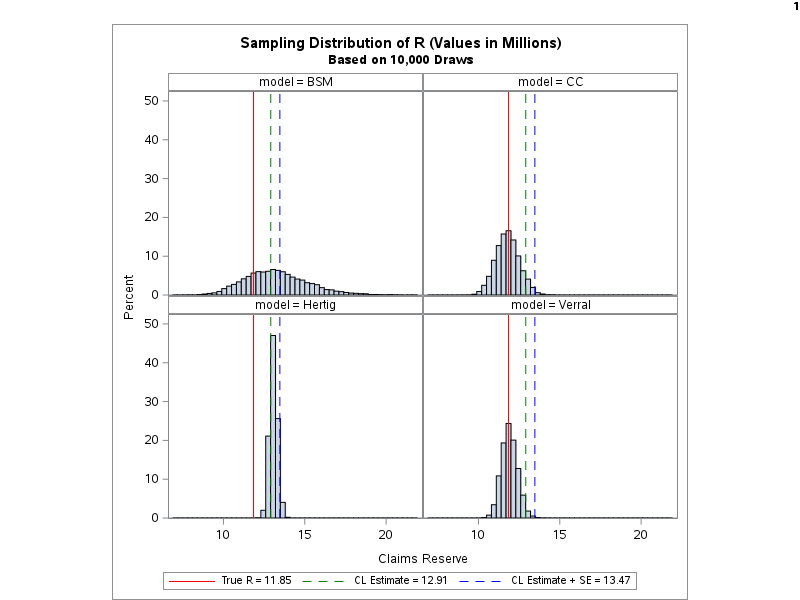}
\end{center}
\label{tab8}
\end{figure}

Based on Table~\ref{chainSum} and Figure~\ref{tab6}, we can say that:
\begin{itemize}
\item The mean and median of $\hat{R}$ for the
CC and Verral models are very close to the true R value (11.85 million), whereas the mean and median
$\hat{R}$ for Hertig and BSM models as well as $\hat{R}$ by the ChainLadder method
exceed the true R value by about a million. 
\item Using the inter-quartile range (the qrange column) as a measure of spread, the sampling
distribution associated with the BSM model is the widest and that associated with the Hertig model
is the most compact.  The coefficient of variation (CV) also points in the same direction.
\end{itemize}
Even clearer picture of these sampling distributions is provided in Figure~\ref{tab7}
and Figure~\ref{tab8}.  The histograms in Figure~\ref{tab7} allow each sampling
distribution to have their own X-axis, whereas in Figure~\ref{tab8} all the histograms
are plotted with X-axis on the same scale.  In each case, vertical reference lines are drawn
to indicate the true R value (a red line at 11.85 million), the ChainLadder point 
estimate (a dashed-green line at 12.9 million), and the ChainLadder point estimate 
plus its standard error (a dashed blue line at 13.5 million).  In summary, these histograms show
that:
\begin{itemize}
\item The true R value is well within the possible values for the histograms associated 
with the CC, Verral, and BSM models.  The true value is far to the left of the histogram
of the Hertig model.
\item The ChainLadder point estimate and the ChainLadder point estimate 
plus its standard error values are within the range of  all the histograms.
\item The histograms associated with the Hertig and Verral models are somewhat compact (possibly too compact
in the case of the Hertig model).
\end{itemize}
In practice, the true R value is unknown. To ensure adequate reserves, a higher percentile of the sampling 
distribution of $\hat{R}$ for a well-fitting model is chosen.  In our illustration, the two well-fitting 
models are CC and Verral.
 By taking the third quartile (Q3) of the sampling distribution of $\hat{R}$ as the suggested reserve, 
the reserve would be around 12.3 million based on either the CC model or the Verral model.
In this illustration, the true R value is 11.85 million. The third quartile (or EstimatePlusSE for ChainLadder) 
for the other models/methods considered (13.2 million for Hertig, 14.5 million for BSM, and 13.5 million for 
ChainLadder) also would have been adequate. However, these reserve values would have been larger than 
the best-fitting model-based reserve suggestions (CC or Verral) by about a million.
\section{Summary}
\label{Summary}
Assuming access to sufficiently feature-rich state space modeling software (such as the CSSM 
procedure in {SAS Viya$^\circledR$}/Econometrics), this article describes a step-by-step 
process for claims-reserve determination based on a given SSM. To highlight common 
statistical issues in real-life claims reserving, the illustration uses a historical claims-table 
that includes both past and future claims, allowing different reserving methods to be 
compared with the ground truth.

The focus is primarily on the steps of the SSM-based claims reservation process. The forms of the SSMs used in 
the illustration are not discussed, but these details are available in the provided references. 
Many other SSMs could have been used for claims reserving.

This article is the first in a series explaining SSM-based claims reserving in practice. The next article will 
summarize the findings of a project testing different model selection strategies for a robust SSM-based 
claims reserving process. The test bed for this project includes 20 complete claims-tables from various 
insurance business categories and a rich selection of SSMs as candidate models.
\section[Disclaim]{Disclaimer}
The views and opinions expressed in this article are solely those of the author 
and do not necessarily reflect the official policy or position of the author's employer. 
The accompanying code is provided "as is," without any warranties, express or implied, including but not 
limited to the implied warranties of merchantability and fitness for a particular purpose. 
The author and the employer shall not be liable for any damages arising from the use of the code.
\appendix
\section[SSMFramework]{SSM Framework and Notation}
\label{SSMFramework}
 All the SSMs discussed in this article are special cases of
the following form:
\begin{eqnarray}
\mathbf{y}_{t} & = & \mathbf{Z}_{t}  \boldsymbol{\alpha}_{t} + 
                                      \mathbf{X}_{t}  \boldsymbol{\beta} + \boldsymbol{\epsilon}_{t}  
                         \;\;\quad  \quad            \quad \quad\text{Observation Equation} \nonumber \\
 \boldsymbol{\alpha}_{t+1}  & = & \mathbf{T}_{t}  \boldsymbol{\alpha}_{t} +
                                                   \mathbf{W}_{t} \boldsymbol{\gamma}  +
                                                              \boldsymbol{\eta}_{t+1}  \quad \quad
                                                              \quad\text{State Equation}  \label{basic} \\
 \boldsymbol{\alpha}_{0}  & = &    \boldsymbol{\eta}_{0} \quad \quad \quad \quad \quad
                              \;\;\quad  \quad                 \quad \quad \; \; \; \;\;\text{Partially Diffuse Initial Condition} \nonumber 
\end{eqnarray}

\begin{itemize}
\item $\mathbf{y}_{t} , t = 1, 2, \cdots$ is a sequence of response vectors.    The number
of responses at different times, i.e., the dimension of $\mathbf{y}_{t}$ at different times, need not be
the same and, some or all elements of $\mathbf{y}_{t}$ can be missing.  In fact, missing
measurements indicate that their values are to be predicted using the remaining
observed data.
\item The observation equation expresses the response vector as 
a sum of three terms: $\mathbf{Z}_{t}  \boldsymbol{\alpha}_{t}$ denotes
the contribution of the state vector $\boldsymbol{\alpha}_{t}$,
$\mathbf{X}_{t}  \boldsymbol{\beta}$ denotes
the contribution of the regression vector $\boldsymbol{\beta}$,
and $\boldsymbol{\epsilon}_{t}$ is a zero-mean, Gaussian noise vector with
diagonal covariance matrix.  The dimension of the state vector, $\boldsymbol{\alpha}_{t}$, does not change
with time.  The design matrices $\mathbf{Z}_{t}$ and $\mathbf{X}_{t}$ are of compatible dimensions.  
\item According to the state equation,  $\boldsymbol{\alpha}_{t+1}$, the state at time $(t+1)$,
is a linear transformation of the previous state, $\boldsymbol{\alpha}_{t}$, 
plus $ \mathbf{W}_{t} \boldsymbol{\gamma}$ (a contribution of regression vector $\boldsymbol{\gamma}$),
plus a random
disturbance, $\boldsymbol{\eta}_{t+1}$, which is a zero-mean, Gaussian vector with
covariance $\mathbf{Q}_{t}$ that need not be diagonal.  
The elements of the state transition matrix $\mathbf{T}_{t}$, the
disturbance covariance $\mathbf{Q}_{t}$, and the design matrix $\mathbf{W}_{t}$
are known.
\item The initial state, $\boldsymbol{\alpha}_{0}$, is assumed to be a Gaussian vector with
known mean, and covariance $\mathbf{Q}_{0}$.  In many cases, no prior information about
some elements of $\boldsymbol{\alpha}_{0}$ is available.   In such cases, their variances 
are taken to be infinite and these elements are called diffuse.
\item The noise vectors in the observation and state equations, $\boldsymbol{\epsilon}_{t}$,
$\boldsymbol{\eta}_{t}$, and the initial condition $\boldsymbol{\alpha}_{0}$, are assumed
to be mutually independent.
\item The elements of system matrices $\mathbf{Z}_{t}, \text{Cov}(\boldsymbol{\epsilon}_{t}), 
\mathbf{T}_{t}, \mathbf{Q}_{t}$, and $\mathbf{Q}_{0}$
are assumed to be completely known, or some of them can be functions of a small set of unknown 
parameters (to be estimated from the data). 
\end{itemize}
The latent vector $\boldsymbol{\alpha}_{t}$ 
can often be partitioned into meaningful sub-blocks (with corresponding blocking
of the design matrix $\mathbf{Z}_{t}$).  In these cases the
observation equation in Equation~\ref{basic} gets the following form:
\[
\mathbf{y}_{t} = \boldsymbol{\mu}_{t} + \boldsymbol{\omega}_{t} + \cdots + 
                              \mathbf{X}_{t}  \boldsymbol{\beta} + \boldsymbol{\epsilon}_{t}  
\]
where the terms $\boldsymbol{\mu}_{t}$, $ \boldsymbol{\omega}_{t}, \cdots$
might represent a time-varying mean-level, a seasonal pattern, and so on.
Such linear combinations of the state sub-blocks are called components.  
When the data from a longitudinal study is assumed to follow an SSM, 
the data analysis is greatly helped by the well-known
(diffuse) Kalman filter, (diffuse) Kalman smoother,
and the simulation smoother algorithms.  Chapters 4, 5, 6, and 7 of \cite{dk}
explain how these algorithms provide the following:
\begin{itemize}
\item Maximum likelihood estimates of the unknown model parameters that are obtained by maximizing
the marginal likelihood.
\item A variety of diagnostic measures for model evaluation.
\item Full-sample estimates of the latent vectors 
$\boldsymbol{\alpha}_{t}$, $\boldsymbol{\beta}$, $\boldsymbol{\gamma}$, and the model components 
such as $\boldsymbol{\mu}_{t}, \boldsymbol{\omega}_{t}, \cdots$, at all time points.
The full-sample estimates are also called the smoothed estimates in the SSM literature.
\item Full-sample predictions of all missing response values.
\item random draws of 
$(\boldsymbol{\beta}, \boldsymbol{\gamma}, \boldsymbol{\alpha}_{1}, \boldsymbol{\alpha}_{2}, \cdots, \boldsymbol{\alpha}_{n})$
from the conditional distribution of 
$(\boldsymbol{\beta}, \boldsymbol{\gamma}, \boldsymbol{\alpha}_{1}, \boldsymbol{\alpha}_{2}, \cdots, \boldsymbol{\alpha}_{n})$ given the full sample $\mathbf{Y} = (\mathbf{y}_{t}, t=1, 2, \cdots, n) $.
\end{itemize}
The documentation of the CSSM procedure contains more precise details about these topics: 
for model-fitting and forecasting see 
\url{https://go.documentation.sas.com/doc/en/pgmsascdc/v_061/casecon/casecon_cssm_details08.htm},
and for simulation smoothing see
\url{https://go.documentation.sas.com/doc/en/pgmsascdc/v_061/casecon/casecon_cssm_details36.htm}.
\bibliography{ibnr}

\begin{thebibliography}{11}
\providecommand{\natexlab}[1]{#1}
\providecommand{\url}[1]{\texttt{#1}}
\expandafter\ifx\csname urlstyle\endcsname\relax
  \providecommand{\doi}[1]{doi: #1}\else
  \providecommand{\doi}{doi: \begingroup \urlstyle{rm}\Url}\fi

\bibitem[Atherino(2010)]{ather}
R.~Atherino.
\newblock A method for modelling varying run-off evolutions in claims
  reserving.
\newblock \emph{ASTIN Bulletin: The Journal of the IAA}, 40\penalty0
  (2):\penalty0 917--946, 2010.

\bibitem[Chukhrova and Johannssen(2021)]{Chukhrova}
N.~Chukhrova and A.~Johannssen.
\newblock Stochastic claims reserving methods with state space representations:
  A review.
\newblock \emph{Risks}, 9, 2021.

\bibitem[De~Jong(2004)]{deJong04}
P.~De~Jong.
\newblock Forecasting general insurance liabilities.
\newblock Technical report, Macquarie University Actuarial Studies Research
  Papers, 2004.

\bibitem[De~Jong and Zehnwirth(1983)]{deJong83}
P.~De~Jong and B.~Zehnwirth.
\newblock Claims reserving, state-space models and the kalman filter.
\newblock \emph{Journal of the Institute of Actuaries}, 4\penalty0
  (1):\penalty0 157--181, 1983.

\bibitem[Durbin and Koopman(2012)]{dk}
J.~Durbin and S.~J. Koopman.
\newblock \emph{Time Series Analysis by State Space Methods, 2nd ed.}
\newblock Oxford University Press, Oxford, 2012.

\bibitem[Gesmann et~al.(2023)Gesmann, Murphy, Zhang, Carrato, Wuthrich,
  Concina, and {Dal Moro}]{CL}
Markus Gesmann, Daniel Murphy, Yanwei~(Wayne) Zhang, Alessandro Carrato, Mario
  Wuthrich, Fabio Concina, and Eric {Dal Moro}.
\newblock \emph{ChainLadder: Statistical Methods and Models for Claims
  Reserving in General Insurance}, 2023.
\newblock URL \url{https://CRAN.R-project.org/package=ChainLadder}.
\newblock R package version 0.2.18.

\bibitem[Hendrych and Cipra(2021)]{Hendrych}
R.~Hendrych and T.~Cipra.
\newblock Applying state space models to stochastic claims reserving.
\newblock \emph{Astin Bulletin}, 51\penalty0 (1), 2021.

\bibitem[Mack(1993)]{mack93}
T.~Mack.
\newblock Distribution-free calculation of the standard error of chain ladder
  reserve estimates.
\newblock \emph{Astin Bulletin}, 23:\penalty0 213--25, 1993.

\bibitem[SAS(2025)]{cssm1}
\emph{The CSSM Procedure: SAS/Econometrics User's Guide}.
\newblock SAS Institute Inc., Cary, NC, 2025.
\newblock URL
  \url{https://go.documentation.sas.com/doc/en/pgmsascdc/v\_061/casecon/casecon\_cssm\_toc.htm}.

\bibitem[Taylor and McGuire(2016)]{Taylor16}
G.~Taylor and G.~McGuire.
\newblock \emph{Stochastic Loss Reserving Using Generalized Linear Models}.
\newblock Casualty Actuarial Society, Arlington, Virginia, 2016.

\bibitem[Verrall(1994)]{verr94}
R.~J. Verrall.
\newblock A method for modelling varying run-off evolutions in claims
  reserving.
\newblock \emph{ASTIN Bulletin: The Journal of the IAA}, 24\penalty0
  (2):\penalty0 325--332, 1994.

\end{thebibliography}
\end{document}